\documentclass[12pt]{iopart}

\usepackage{graphicx}
\usepackage{multirow}
\usepackage{dcolumn}
\usepackage{cite}

\begin{document}

\title[Rayleigh-Ritz method]{Rayleigh-Ritz variational method with suitable asymptotic behaviour}
\author{Javier Garcia \ and Francisco M Fern\'andez}

\address{INIFTA (UNLP, CCT La Plata-CONICET), Divisi\'on Qu\'imica Te\'orica,
Blvd. 113 S/N,  Sucursal 4, Casilla de Correo 16, 1900 La Plata,
Argentina}

\ead{fernande@quimica.unlp.edu.ar}

\maketitle

\begin{abstract}
We discuss Rayleigh-Ritz variational calculations with nonorthogonal basis
sets that exhibit the correct asymptotic behaviour. We construct the
suitable basis sets for general one-dimensional models and illustrate the
application of the approach on two double-well oscillators proposed recently
by other authors. The rate of convergence of the variational method proves
to be considerably greater than the one exhibited by the recently developed
orthogonal polynomial projection quantization.
\end{abstract}

\section{Introduction}

\label{sec:intro}

In a recent paper Handy and Vrinceanu\cite{HV13} proposed a method
for the calculation of energy eigenvalues that is based on the
projection of the bound-state wavefunction onto sets of orthogonal
polynomials. The approach named orthogonal polynomial projection
quantization (OPPQ) proved to be rapidly converging and more
stable than the Hill determinant methods.

Among other models the authors considered the sextic $V_{S}(x)=x^{6}-4x^{2}$
and quartic $V_{Q}(x)=x^{4}-5x^{2}$ two-well oscillators. The bound states
behave asymptotically as $\psi (x)\sim e^{-x^{4}/4}$ and $\psi (x)\sim
e^{-|x|^{3}/3}$ in the former and latter case, respectively. Handy and
Vrinceanu\cite{HV13} chose the reference functions $R_{G}(x)=e^{-x^{2}/2}$
and $R_{TT}(x)=e^{-x^{4}/4}$ and showed that the latter is preferable for $%
V_{S}(x)$ while the former is more convenient for $V_{Q}(x)$. In fact, $%
R_{TT}(x)$ exhibits the correct asymptotic behaviour for the potential $%
V_{S}(x)$. At first sight it appears to be surprising that the
authors did not try the reference function $R(x)=e^{-|x|^{3}/3}$,
which is expected to be suitable for the quartic double well,
since their approach permits the use of arbitrary nonanalytic
positive reference functions\cite{HV13}.

The purpose of this paper is to show that it is quite straightforward to
apply the variational Rayleigh-Ritz method (RRM) with a basis set that
exhibits the correct asymptotic behaviour of the eigenfunctions for the
models discussed above. In addition to it, we deem it worthwhile to compare
the well known, extremely reliable and widely used RRM with the recently
developed OPPQ.

In section~\ref{sec:basis_set} we develop the basis sets with suitable
asymptotic behaviours for some general one-dimensional models. In section~%
\ref{sec:results} we calculate the eigenvalues for the oscillators $V_{S}(x)$
and $V_{Q}(x)$ with three basis sets having different asymptotic behaviours,
including the correct one for each model. We also compare RRM and OPPQ
results for two models and two basis sets. Finally, in section~\ref
{sec:conclusions} we summarize the main conclusions of the paper.

\section{Basis functions with suitable asymptotic behaviour}

\label{sec:basis_set}

In this section we show how to build a non-orthogonal basis set with the
appropriate asymptotic behaviour at infinity. To this end we generalize a
procedure proposed recently by Fern\'{a}ndez\cite{F13} for a particular
case. For simplicity we focus on the one-dimensional eigenvalue equation
\begin{equation}
-\psi ^{\prime \prime }(x)+V(x)\psi (x)=E\psi (x),  \label{eq:Schro}
\end{equation}
and assume tat
\begin{equation}
\lim\limits_{|x|\rightarrow \infty }x^{-2k}V(x)=a>0.  \label{eq:lim_V}
\end{equation}
Under such condition the eigenfunctions behaves asymptotically as
\begin{eqnarray}
\psi (x) &\sim &e^{-|S_{k}(x)|}  \nonumber \\
S_{k}(x) &=&\frac{\sqrt{a}}{k+1}x^{k+1}.  \label{eq:Psi_asymp}
\end{eqnarray}
We consider the following cases:

Case 1: Parity-invariant potential $V(-x)=V(x)$.

a) $k$ even. The non-orthogonal basis set is of the form
\begin{eqnarray}
f_{j}(x) &=&|x|^{j}e^{-|S_{k}(x)|},\,j=0,2,3,\ldots \mathrm{\ even\,states},
\nonumber \\
f_{j}(x) &=&x|x|^{j}e^{-|S_{k}(x)|},\,j=0,1,\ldots \mathrm{\ odd\,states},
\label{eq:basis_k_even}
\end{eqnarray}

b) $k$ odd. In this case we choose
\begin{eqnarray}
f_{j}(x) &=&x^{2j+s}e^{-S_{k}(x)},\,j=0,1,\ldots  \nonumber \\
s &=&\left\{
\begin{array}{c}
0\mathrm{\ }\mathrm{even\,states} \\
1\mathrm{\ }\mathrm{odd\,states}
\end{array}
\right. .  \label{eq:basis_k_odd}
\end{eqnarray}

Case 2: Asymmetric potential $V(-x)\neq V(x)$. The basis set is
\begin{equation}
f_{j}(x)=x^{j}e^{-|S_{k}(x)|},\,j=0,1,\ldots
\label{eq:basis_V_assym}
\end{equation}

For concreteness in what follows we consider two of the examples discussed
by Handy and Vrinceanu\cite{HV13}
\begin{equation}
V_{Q}(x)=x^{4}-5x^{2},  \label{eq:V_Q}
\end{equation}
and
\begin{equation}
V_{S}(x)=x^{6}-4x^{2},  \label{eq:V_S}
\end{equation}
with asymptotic behaviours given by $S_{2}(x)=x^{3}/3$ and $S_{3}(x)=x^{4}/4$%
, respectively. However, they chose reference functions associated to $%
S_{3}(x)$ and $S_{1}(x)=x^{2}/2$ for the two models.

The RRM enables us to obtain the eigenvalues approximately from the roots of
the secular determinant
\begin{equation}
\left| \mathbf{H}-E\mathbf{S}\right| =0,  \label{eq:sec_det}
\end{equation}
where $\mathbf{H}$ and $\mathbf{S}$ are $N\times N$ square matrices with
elements
\begin{equation}
H_{ij}=\left\langle f_{i}\right| \hat{H}\left| f_{j}\right\rangle
,\,S_{ij}=\left\langle f_{i}\right| \left. f_{j}\right\rangle ,
\label{eq:mat_el}
\end{equation}
and $\hat{H}=\hat{p}^{2}+V(x)$. Once we have the approximate eigenvalues we
obtain the eigenvectors from the secular equation
\begin{equation}
\left( \mathbf{H}-E\mathbf{S}\right) \mathbf{C=0,}  \label{eq:sec_eq}
\end{equation}
where $\mathbf{C}$ is an $N\times 1$ column matrix with the coefficients $%
c_{j}$ of the variational trial function.

The application of this approach to the Schr\"{o}dinger equation with a
parity-invariant potential is particularly simple because we can restrict
the calculation of the matrix elements to the half line $x>0$\cite{F13}.
Since all the matrix elements reduce to integrals of the form
\begin{equation}
\left\langle f\right| \left. g\right\rangle =\int_{0}^{\infty }f(x)g(x)\,dx,
\label{eq:inner_prod}
\end{equation}
then we do not have to take into account the absolute value of the
coordinate explicitly when $k$ is even.

\section{Results}

\label{sec:results}

\begin{figure}[tbp]
\begin{center}
\bigskip\bigskip\bigskip \includegraphics[width=6cm]{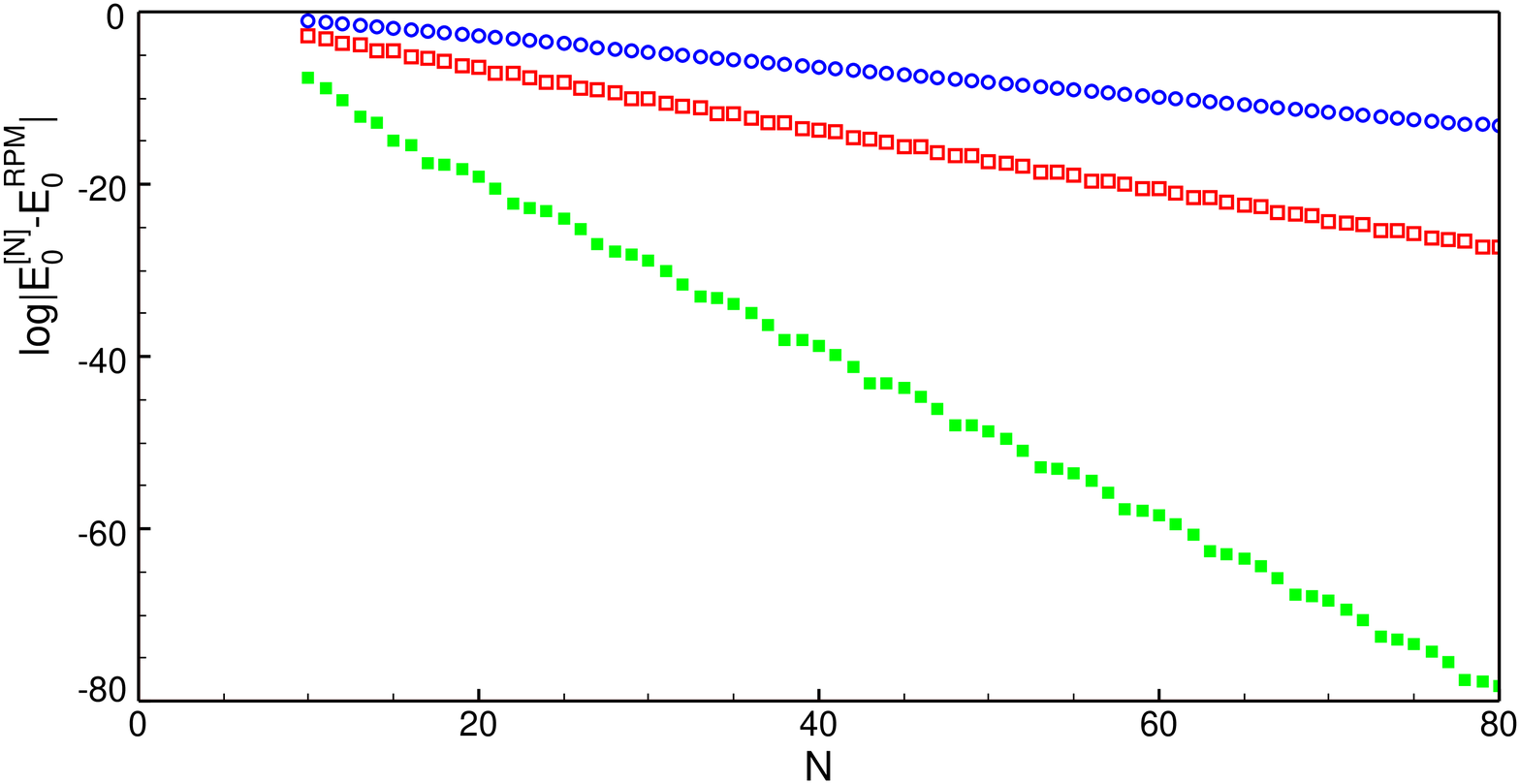} %
\includegraphics[width=6cm]{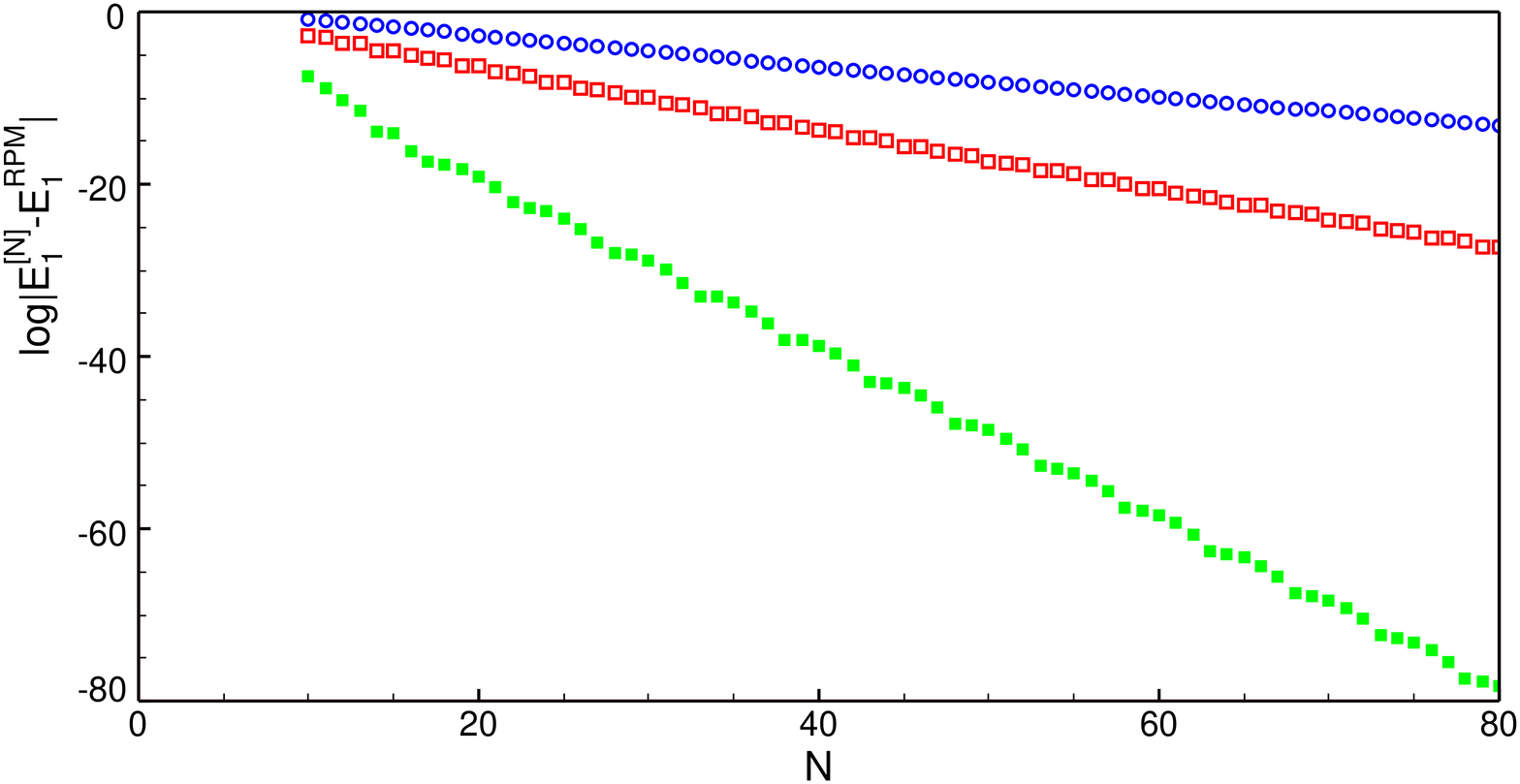} %
\includegraphics[width=6cm]{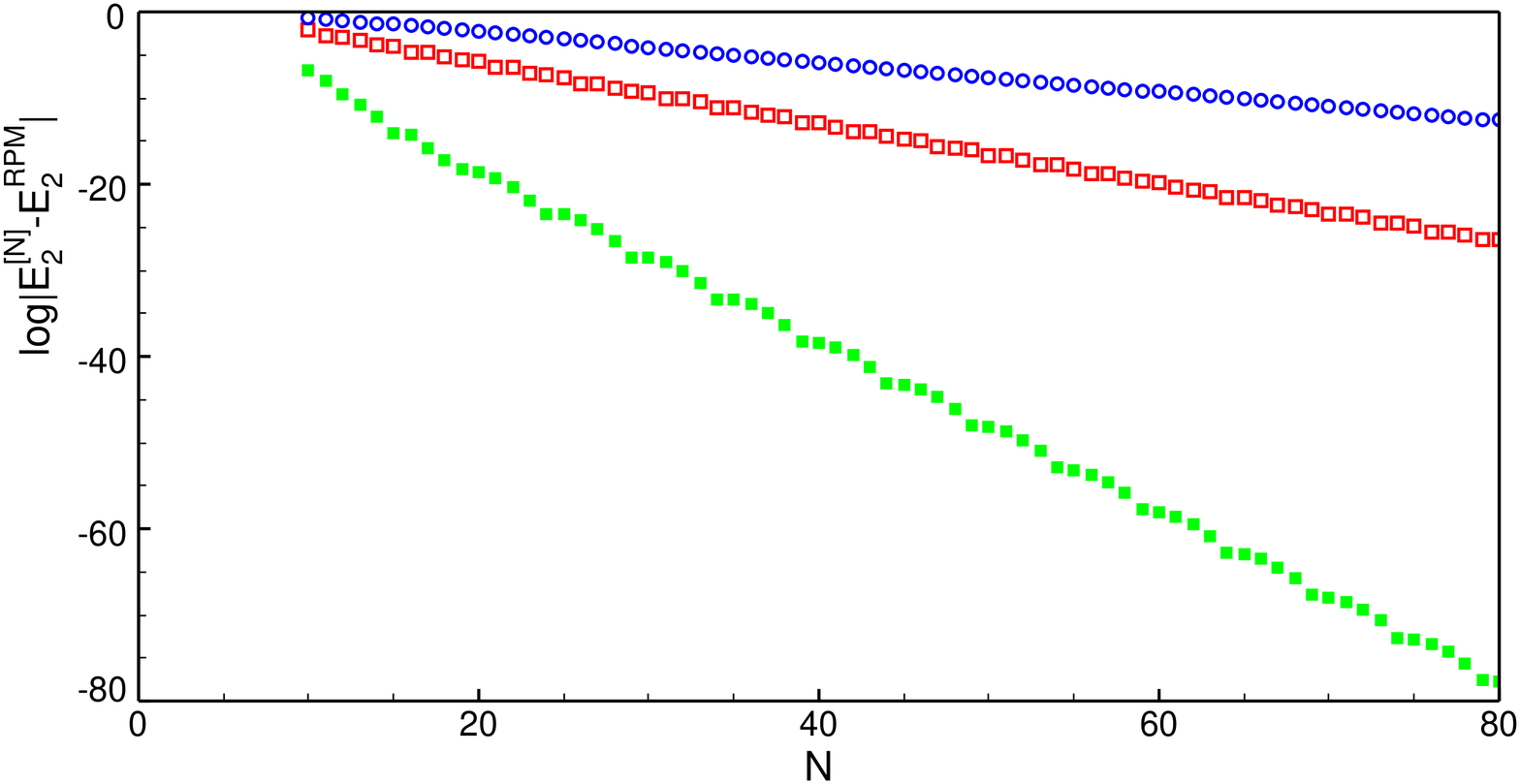} %
\includegraphics[width=6cm]{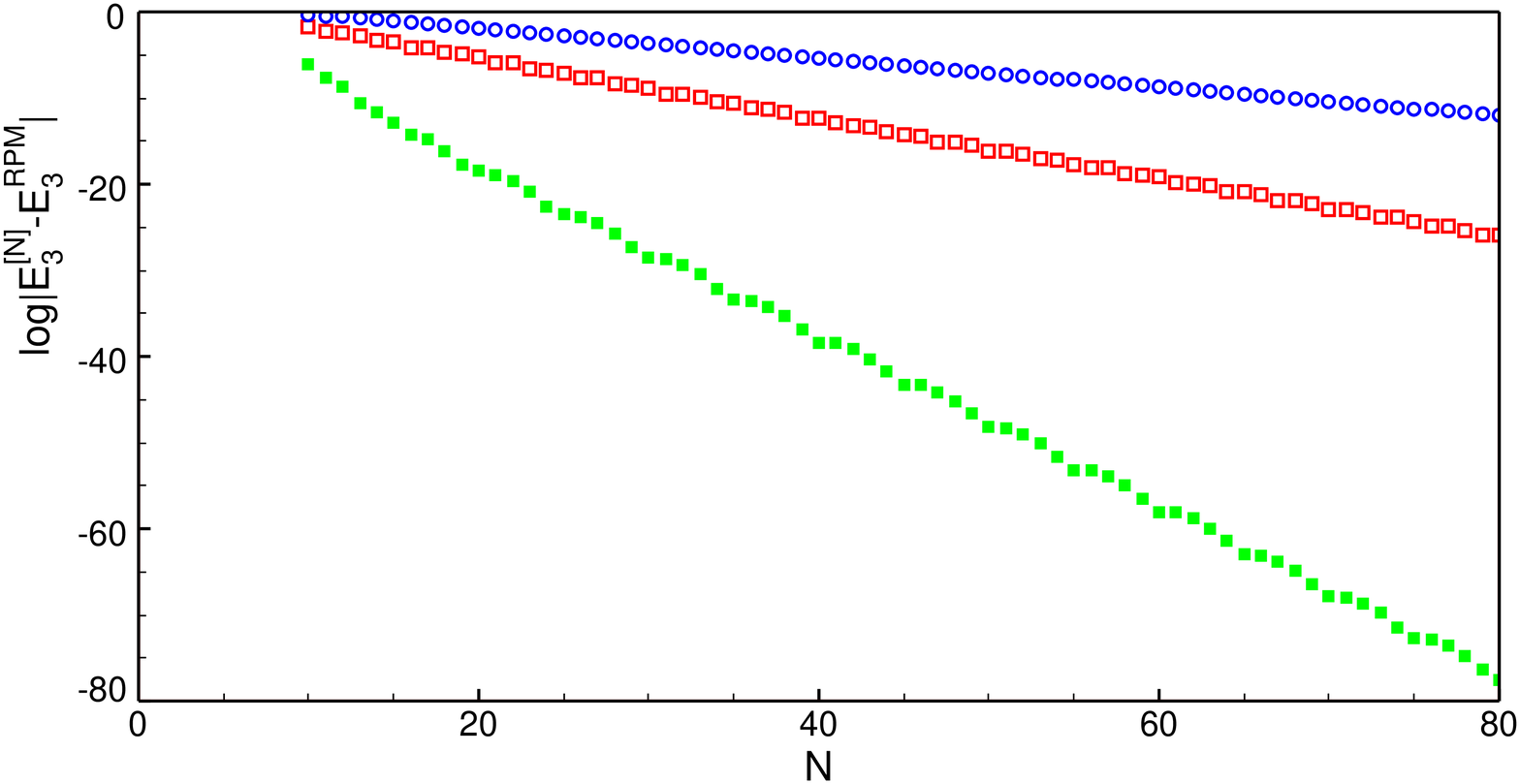}
\end{center}
\caption{Logarithmic errors for the first four eigenvalues of the double
well $V_Q(x)$ calculated by means of the basis sets with $S_1(x)$ (squares,
red) $S_2(x)$ (filled squares, green) and $S_3(x)$ (circles, blue)}
\label{fig:x4x2}
\end{figure}

\begin{figure}[tbp]
\begin{center}
\bigskip\bigskip\bigskip \includegraphics[width=6cm]{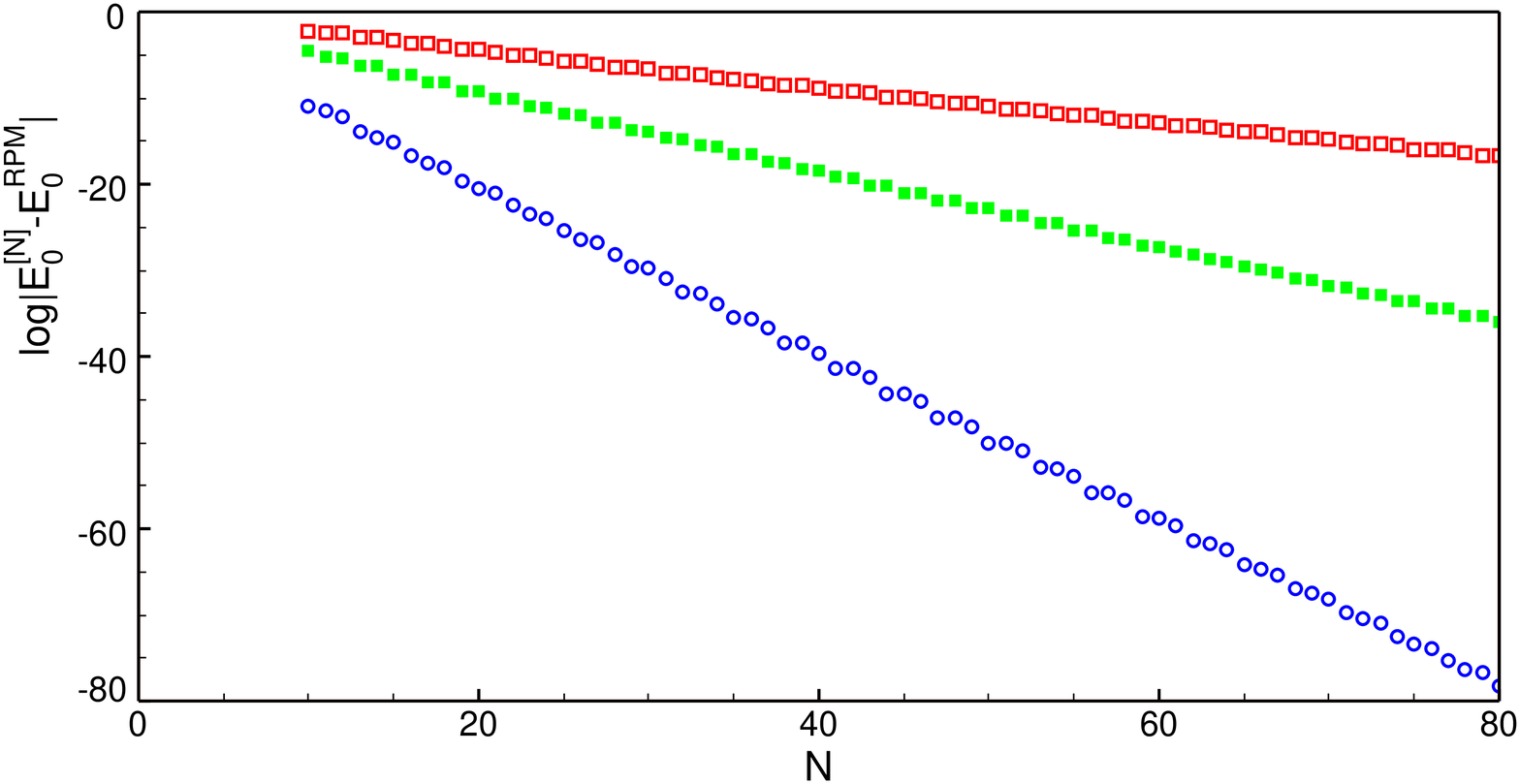} %
\includegraphics[width=6cm]{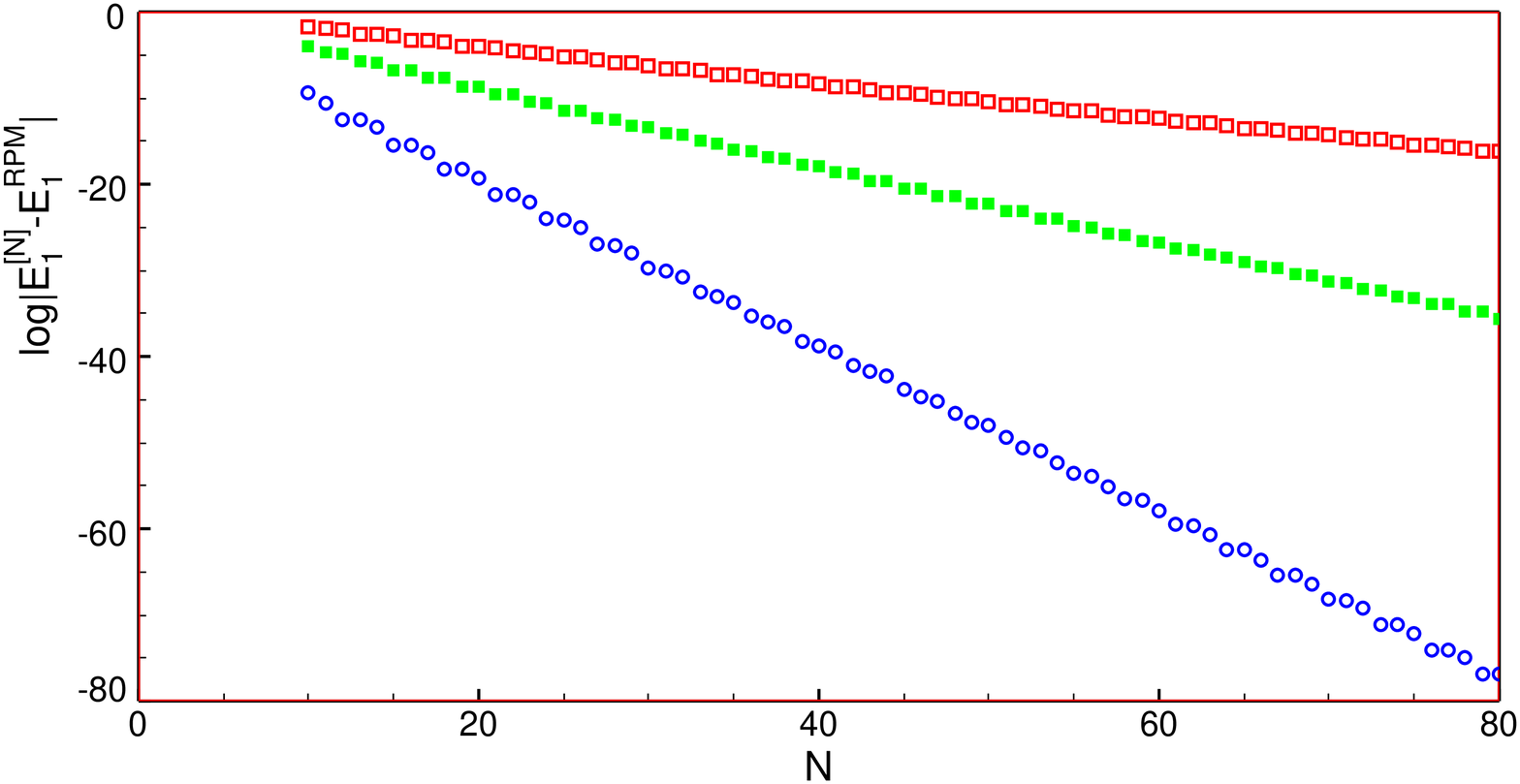} %
\includegraphics[width=6cm]{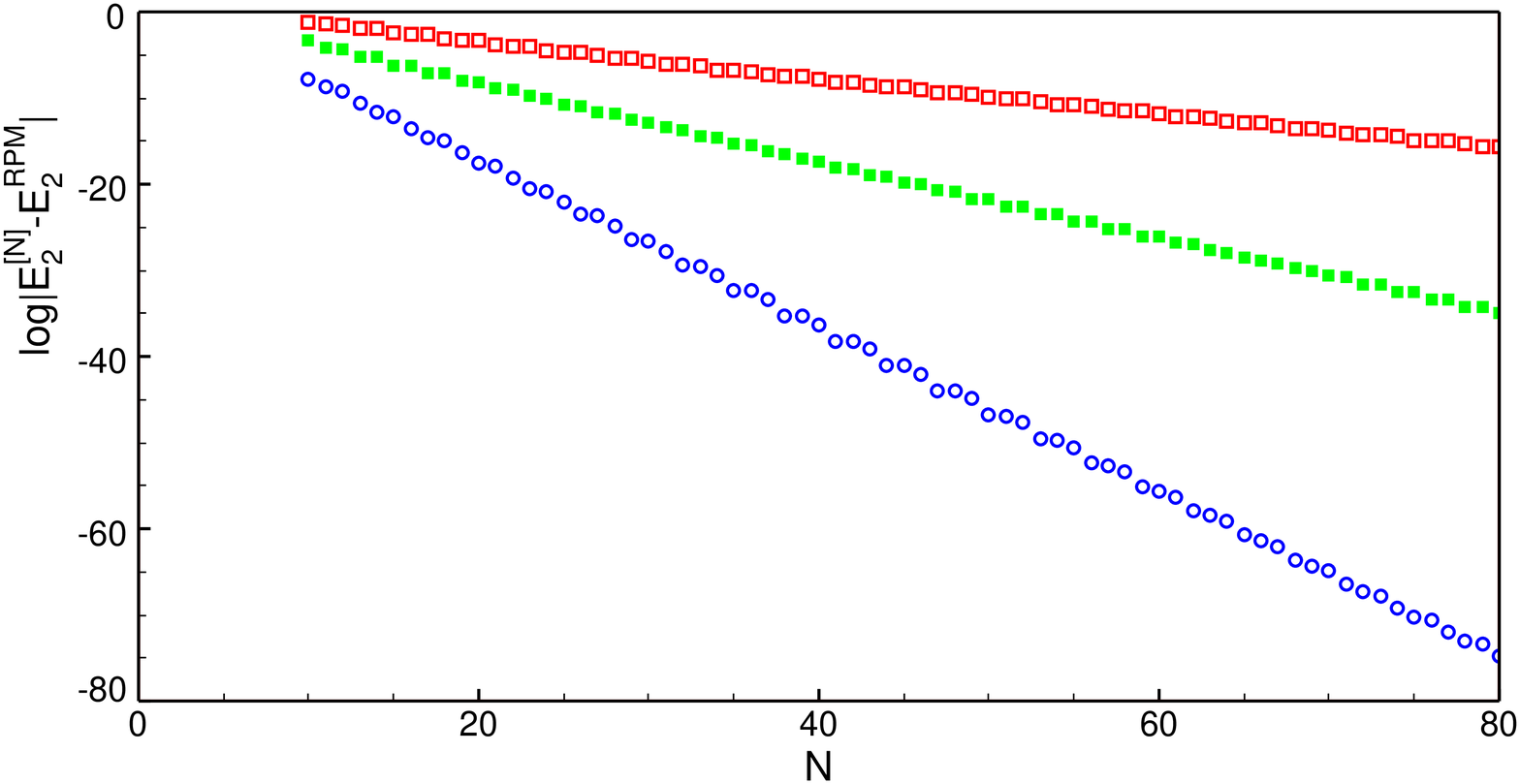} %
\includegraphics[width=6cm]{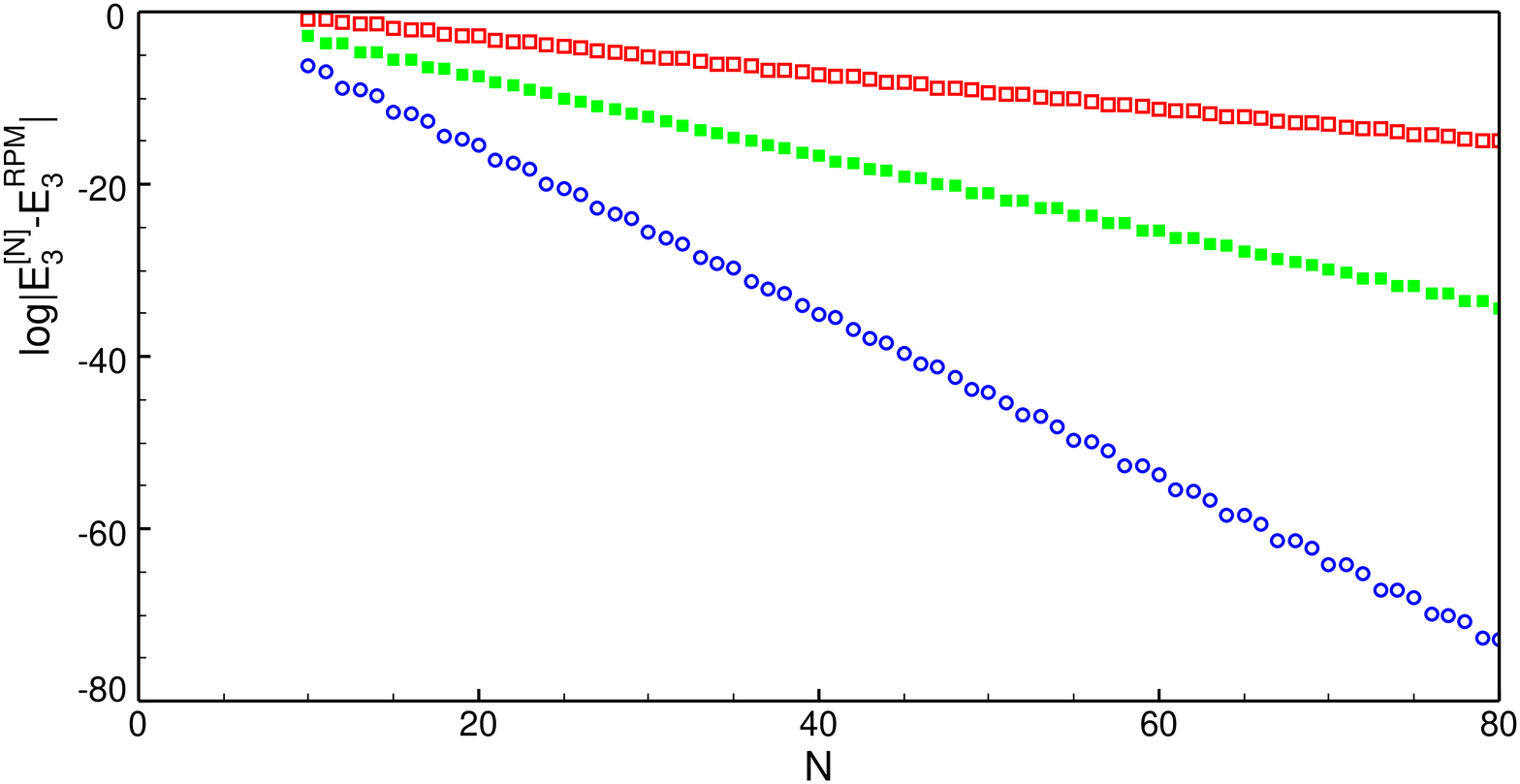}
\end{center}
\caption{Logarithmic errors for the first four eigenvalues of the double
well $V_S(x)$ calculated by means of the basis sets with $S_1(x)$ (squares,
red) $S_2(x)$ (filled squares, green) and $S_3(x)$ (circles, blue)}
\label{fig:x6x2}
\end{figure}

We first verify the effect of the asymptotic behaviour of the basis sets (%
\ref{eq:basis_k_even}) and (\ref{eq:basis_k_odd}) on the rate of convergence
of the RRM. A reasonable estimate of the rate of convergence is the
logarithmic error $L_{N}=\log \left| E_{n}^{(app)}-E_{n}^{(RPM)}\right| $
where $E_{n}^{(app)}$ is the eigenvalue calculated by any of the methods
described in this paper and $E_{n}^{(RPM)}$ is a very accurate result
obtained by means of the RPM\cite{FMT89a,FMT89b}. Figure~\ref{fig:x4x2}
shows $L_{N}$ for the first four eigenvalues of (\ref{eq:V_Q}) calculated by
means of the RRM with the functions $S_{1}(x)$, $S_{2}(x)$ and $S_{3}(x)$ in
terms of the number of basis functions $N$. We see that the rate of
convergence decreases according to $S_{2}(x)>S_{1}(x)>S_{3}(x)$; that is to
say, the RRM converges more rapidly when choosing the correct asymptotic
behaviour $S_{2}(x)$. On the other hand, figure~\ref{fig:x6x2} shows that
the relative rate of convergence of the RRM for the potential (\ref{eq:V_S})
is $S_{3}(x)>S_{2}(x)>S_{1}(x)$. Once again the greater rate of convergence
is given by the correct asymptotic behaviour $S_{3}(x)$. Besides, the second
inequality appears to be reasonable because $S_{2}(x)$ is closer to the
correct asymptotic behaviour than $S_{1}(x)$.

\begin{figure}[tbp]
\begin{center}
\bigskip\bigskip\bigskip %
\includegraphics[width=6cm]{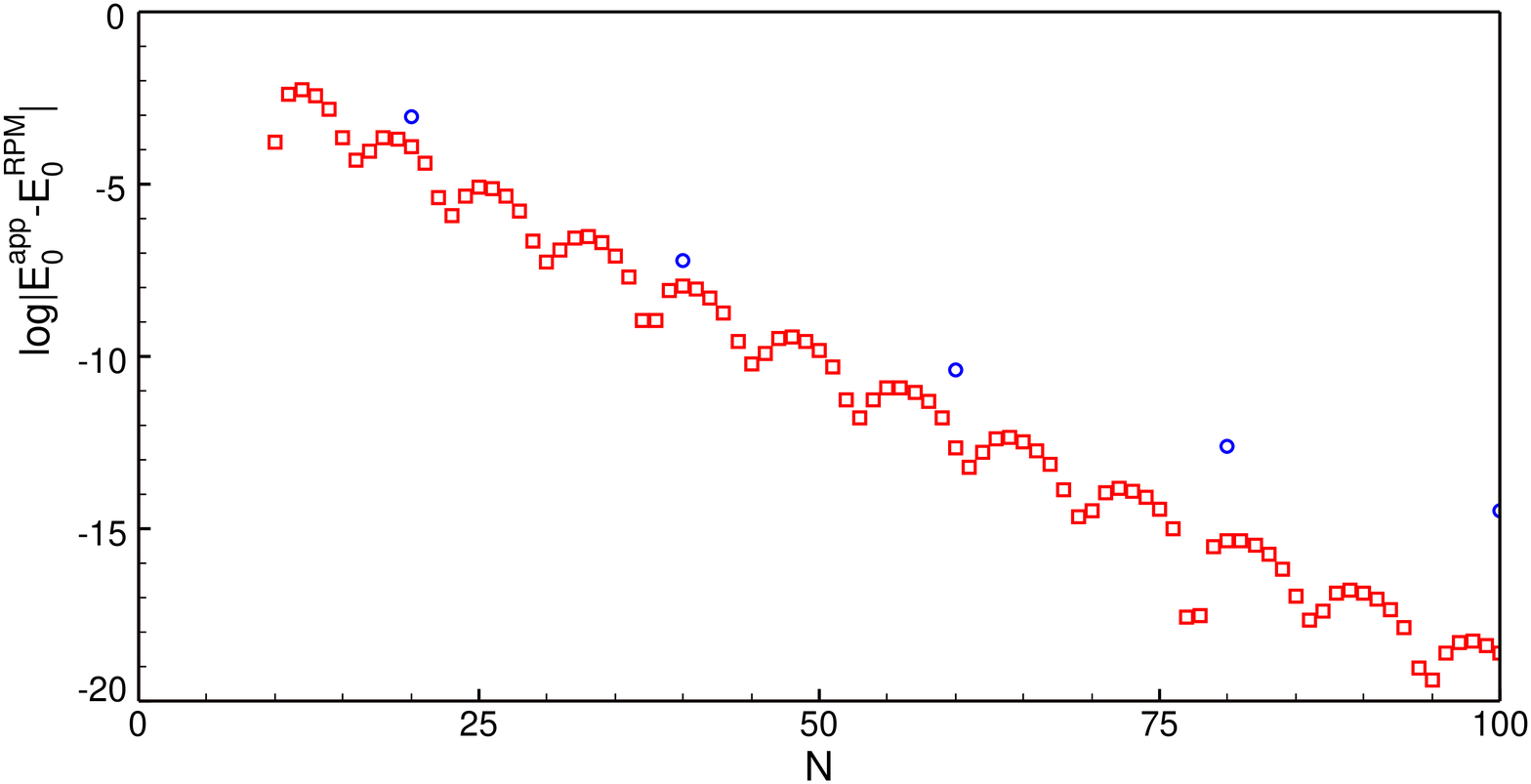} %
\includegraphics[width=6cm]{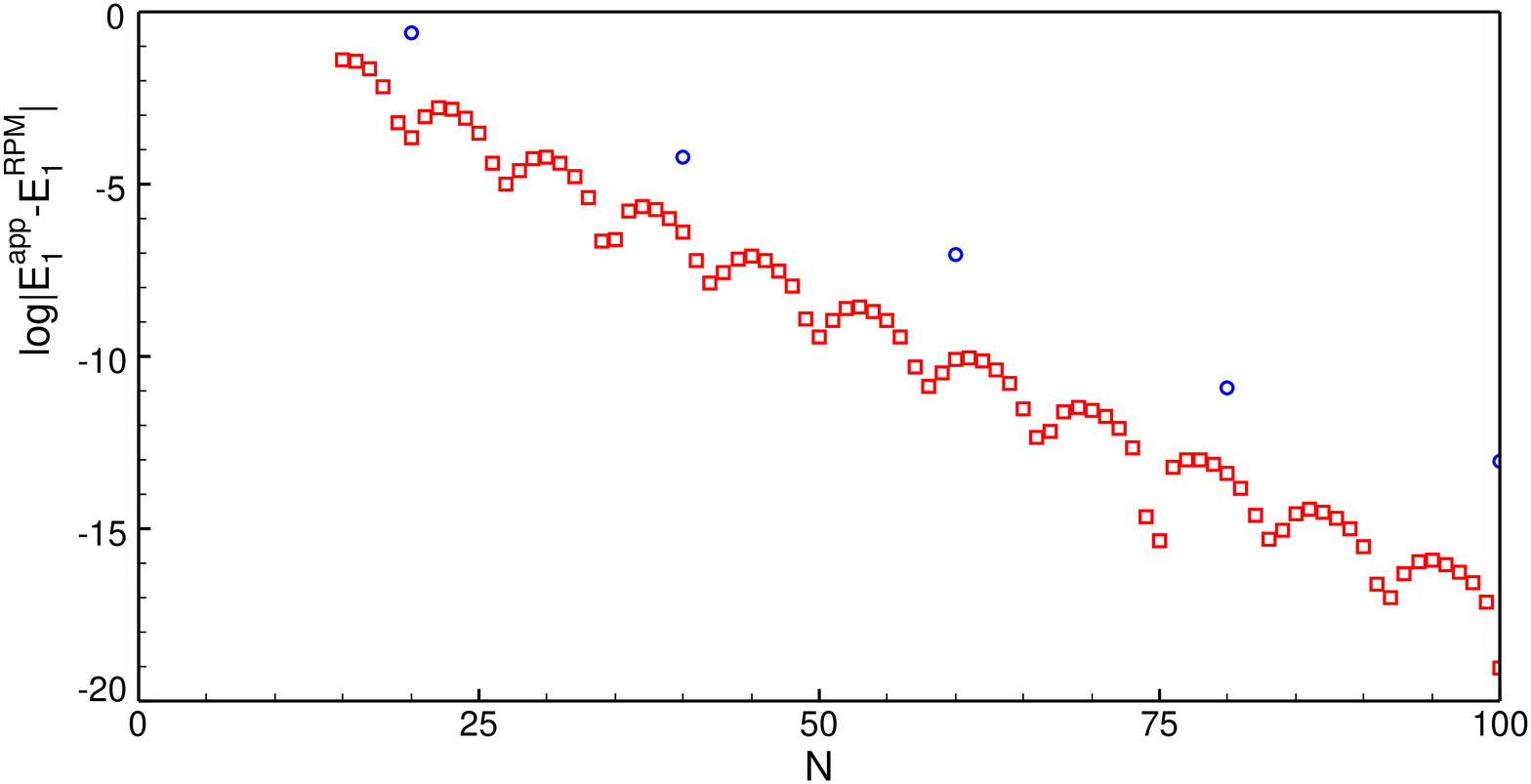} %
\includegraphics[width=6cm]{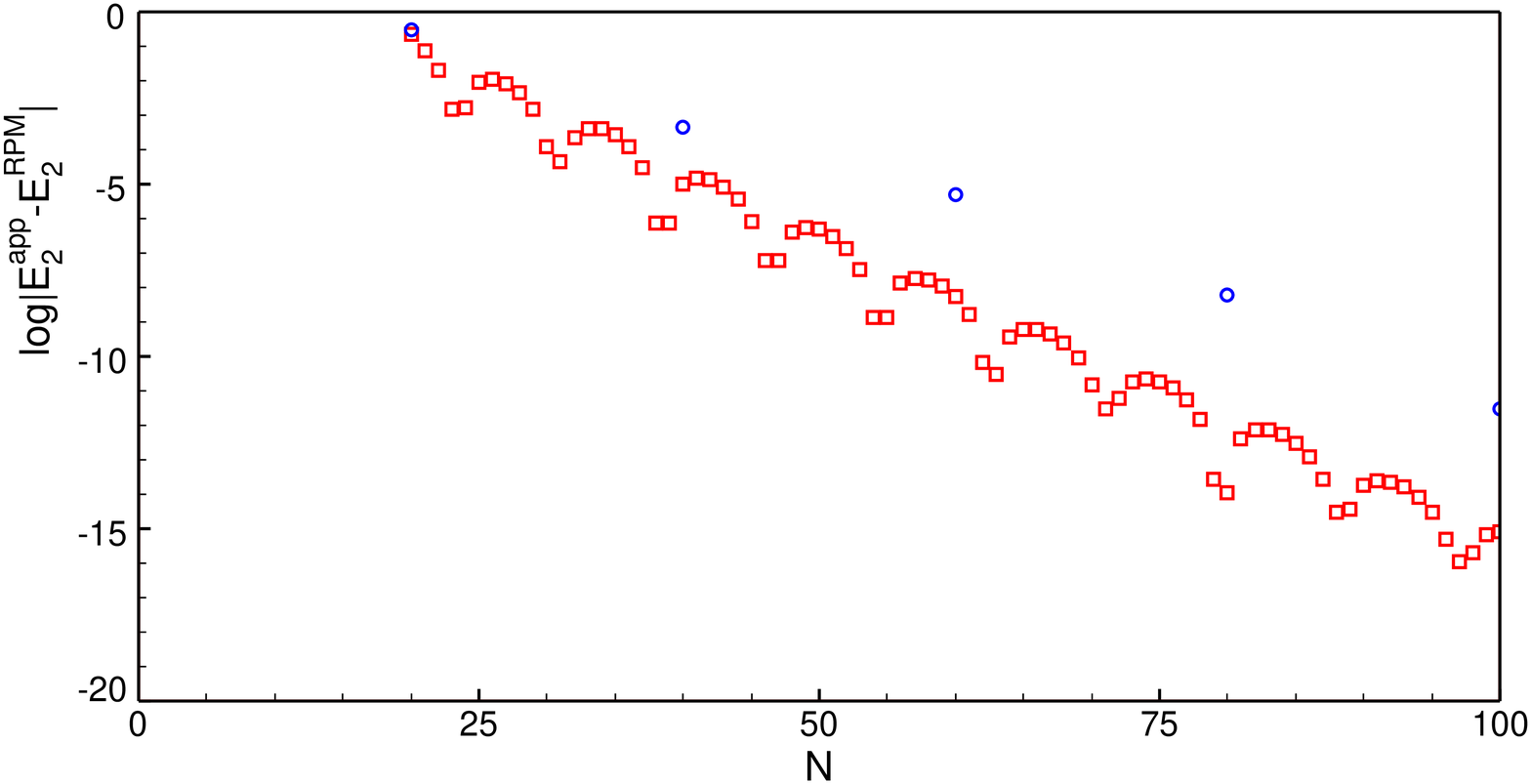} %
\includegraphics[width=6cm]{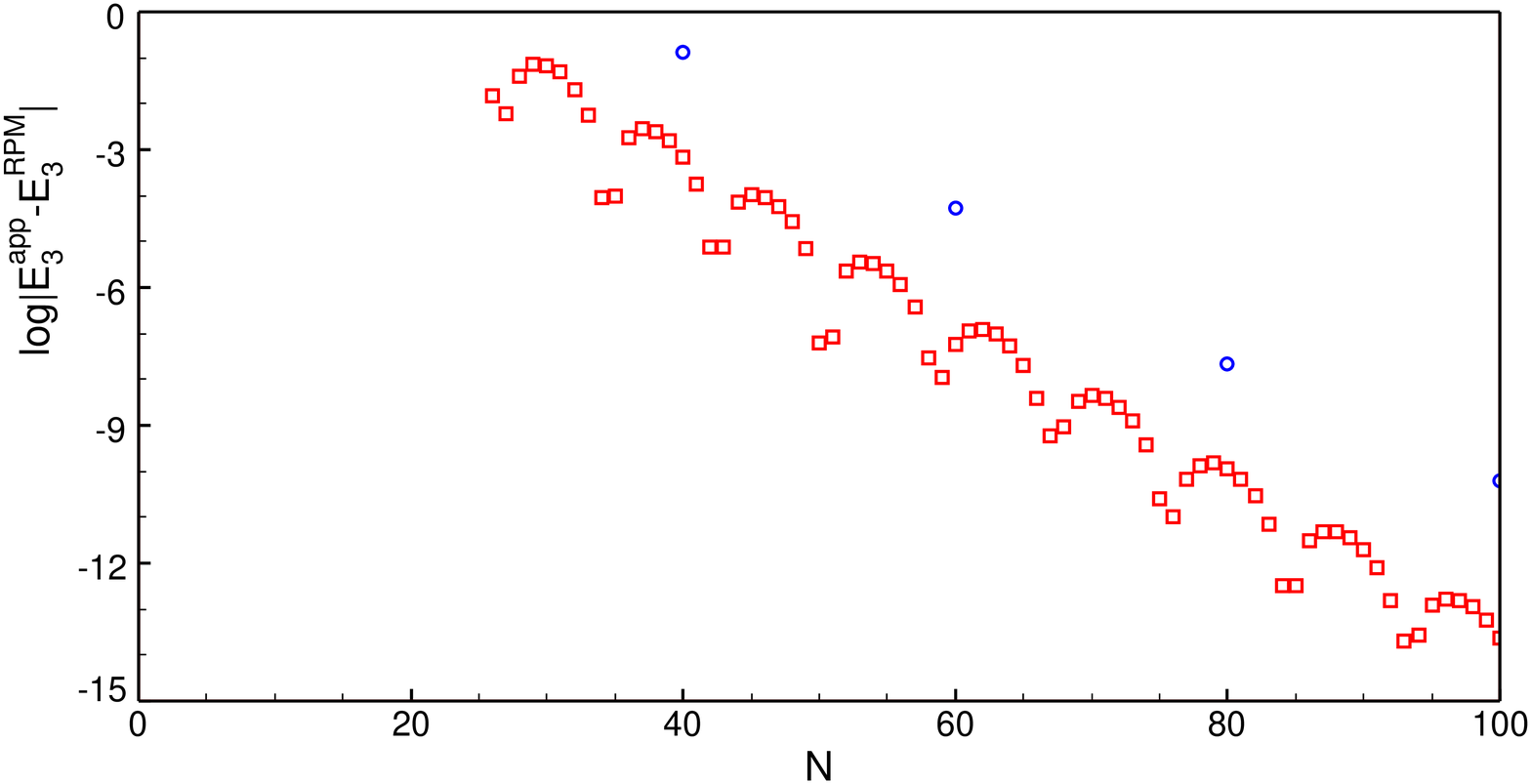} %
\par
\end{center}
\caption{Logarithmic errors for the first four eigenvalues of the
PT-symmetric potential (\ref{eq:ix^3}) calculated by means of the
RRM (squares, red) and the OPPQ method (circles, blue) }
\label{fig:ix3}
\end{figure}

We think that it is also worthwhile to compare the rate of convergence of
the simple, well known and reliable RRM and the rather more elaborate OPPQ
using the same basis set in both approaches. Handy and Vrinceanu\cite{HV13}
chose the reference function $R_{G}(x)=e^{-x^{2}/2}$ for the PT-symmetric
potential
\begin{equation}
V(x)=ix^{3}  \label{eq:ix^3}
\end{equation}
and we therefore choose the function $S_{1}(x)$ for the RRM. More
precisely, instead of the non-orthogonal basis set
(\ref{eq:basis_k_odd}) we resorted to the eigenfunctions of the
harmonic oscillator that are truly consistent with the orthogonal
Hermite polynomials used by those authors. Figure~\ref {fig:ix3}
shows $L_{N}$ for the first four eigenvalues calculated by both
methods. It clearly shows that the RRM rate of convergence is
noticeably greater that the OPPQ one.

\begin{figure}[tbp]
\begin{center}
\bigskip\bigskip\bigskip \includegraphics[width=12cm]{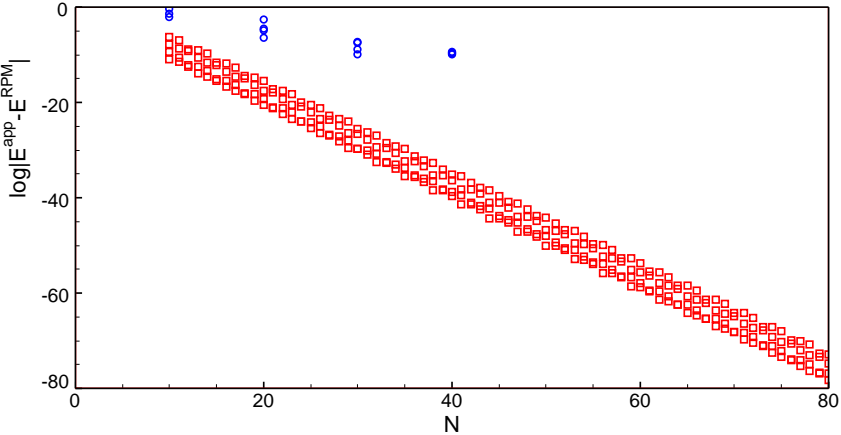}
\par
\end{center}
\caption{Logarithmic errors for the first four eigenvalues of the double
well $V_S(x)$ calculated by means of the RRM (squares, red) and the OPPQ
method (circles, blue)}
\label{fig:x4x2_b}
\end{figure}

The better performance of the RRM is not restricted to the
PT-symmetric cubic oscillator; this approach converges more
rapidly for the two other models discussed above. For example,
figure~\ref{fig:x4x2_b} compares the logarithmic errors of both
approaches for the eigenvalues of the double-well oscillator
(\ref{eq:V_S}). In this case we have chosen the basis set with the
correct asymptotic behaviour $S_{3}(x)$ for the RRM and the OPPQ
results for $R_{TT}(x)=e^{-x^{4}/4}$\cite{HV13}. The difference
between the convergence rates of both approaches is even more
dramatic for this oscillator. However, it makes more sense for
small $N$ because the number of significant digits of the OPPQ
results reported by the authors is rather small for a fair
comparison at large $N$.

Finally, we deem it worthwhile to show the RPM eigenvalues chosen as a
reference for the calculation of the logarithmic errors. We have
\begin{eqnarray}
E_{0} &=&-3.41014276123982947529770965352190919871233904756  \nonumber \\
&&4881868937911775329611301715294  \nonumber \\
E_{1} &=&-3.250675362289235980228513775547736877154601147639  \nonumber \\
&&4241429953014335680690809034749688022953825298  \nonumber \\
E_{2} &=&0.6389195637838381244910101033325042648524013290581  \nonumber \\
&&37207433367771840730088316019330941500824  \nonumber \\
E_{3} &=&2.5812162706174514809779380656962090234197947974759  \nonumber \\
&&598949291704975284539346710703866627200928172
\end{eqnarray}
for (\ref{eq:V_Q})
\begin{eqnarray}
E_{0} &=&-0.5232686221275522394161694971907840611656342225187  \nonumber \\
&&11069953854385633821213450649003542309  \nonumber \\
E_{1} &=&1.00576834022554481670604083074777604686886504417542  \nonumber \\
&&730471341100873617568288708176003637  \nonumber \\
E_{2} &=&5.37497000884004499406051476941823532582175431150133  \nonumber \\
&&8177585996687355671683247232390293  \nonumber \\
E_{3} &=&10.5725850445859121139060615553140114648422138800575  \nonumber \\
&&29217715660995992776130576146017312
\end{eqnarray}
for (\ref{eq:V_S}) and
\begin{eqnarray}
E_{0} &=&1.156267071988113293799219177999951  \nonumber \\
E_{1} &=&4.1092287528096515358436684785613  \nonumber \\
E_{2} &=&7.5622738549788280413518091106314827208  \nonumber \\
E_{3} &=&11.314421820195804402233783948426989
\end{eqnarray}
for (\ref{eq:ix^3}). These quite accurate results may be used as benchmarks
for other approaches.

\section{Conclusions}

\label{sec:conclusions}

We have shown that it is not difficult to introduce the correct
asymptotic behaviour of the wavefunction into the RRM variational
trial function, specially if the potential is parity invariant.
Present results clearly show that the correct asymptotic behaviour
increases the rate of convergence of the approach dramatically. In
principle, the same strategy can be implemented through the
appropriate OPPQ reference function but it has not yet been tried
for the case of $k$ even\cite{HV13}.

We have also shown that the rate of convergence of the RRM is considerably
greater than that for the OPPQ. In addition to it the former approach is
simpler and more straightforward. The integrals that appear in both
approaches are basically the same and can in principle be calculated by the
same algorithms. Here we just compared the results for two oscillators and
two basis sets with different asymptotic behaviours but the trend is exactly
the same for the other possible combinations of model and basis set.

We do not know the reason why the RRM rate of convergence is so much greater
than the OPPQ one. What we already know is that in the case of the Hermitian
Hamiltonians the former approach exhibits the additional advantage that its
approximate eigenvalues tend to the exact ones from above. On the other
hand, the OPPQ eigenvalues do not appear to exhibit any bounding property.

\end{document}